# Oxygen fractionation in dense molecular clouds

*Jean-Christophe Loison[1]\*, Valentine Wakelam[2], Pierre Gratier[2], Kevin M. Hickson[1], Aurore Bacmann[3], Marcelino Agùndez[4], Nuria Marcelino[4], José Cernicharo[4], Viviana Guzman[5], Maryvonne Gerin[6], Javier R. Goicoechea[7], Evelyne Roueff[8], Franck Le Petit[8], Jérome Pety[9,6], Asunción Fuente[10], Pablo Riviere-Marichalar[4]*

\*Corresponding author: <u>jean-christophe.loison@u-bordeaux.fr</u>

[1] *Institut des Sciences Moléculaires (ISM), CNRS, Univ. Bordeaux, 351 cours de la Libération, 33400, Talence, France*
[2] *Laboratoire d'astrophysique de Bordeaux, CNRS, Univ. Bordeaux, B18N, allée Geoffroy Saint-Hilaire, 33615 Pessac, France.*
[3] *Institut de Planétologie et d'Astrophysique de Grenoble (IPAG) UMR 5274, UJF-Grenoble 1 / CNRS-INSU, 38041 Grenoble, France*
[4] *Instituto de Física Fundamental, CSIC, C/ Serrano 123, 28006 Madrid, Spain*
[5] *Joint ALMA Observatory (JAO), Alonso de Córdova 3107, Vitacura, Santiago de Chile, Chile*
[6] *LERMA, Observatoire de Paris, PSL Research University, CNRS, Sorbonne Universités, UPMC Univ. Paris 06, Ecole Normale Supérieure, F-75005 Paris, France*
[7] *Instituto de Física Fundamental, CSIC, Calle Serrano, 121, 28006 Madrid, Spain*
[8] *LERMA, Observatoire de Paris, PSL Research University, CNRS, Sorbonne Universités, UPMC Univ. Paris 06, F-92190 Meudon, France*
[9] *Institut de Radioastronomie Millimétrique (IRAM), 300 rue de la Piscine, 38406 Saint Martin d'Hyères, France*
[10] *Observatorio Astronómico Nacional (OAN, IGN), Apdo 112, E-28803 Alcalá de Henares, Spain*

**ABSTRACT**

We have developed the first gas-grain chemical model for oxygen fractionation (also including sulphur fractionation) in dense molecular clouds, demonstrating that gas-phase chemistry generates variable oxygen fractionation levels, with a particularly strong effect for NO, SO, $O_2$, and $SO_2$. This large effect is due to the efficiency of the neutral $^{18}O + NO$, $^{18}O + SO$, and $^{18}O + O_2$ exchange reactions. The modeling results were compared to new and existing observed isotopic ratios in a selection of cold cores. The good agreement between model and observations requires that the gas-phase abundance of neutral oxygen atoms is large in the observed regions. The $S^{16}O/S^{18}O$ ratio is predicted to vary substantially over time showing that it can be used as a sensitive chemical proxy for matter evolution in dense molecular clouds.

# 1 INTRODUCTION

To understand the transformation of matter from gas and dust present in the interstellar medium to the formation of planetary systems, a precise understanding of the nature and abundance of the various species present prior to cloud collapse is required. With the exception of CO and minor species in the gas phase and a few species on interstellar ices, the exact chemical composition of molecular clouds is still unknown. This is due, in particular, to the difficulty to detect species without a dipole moment (such as $O_2$ and $N_2$). It is unknown, for example, what fraction of elemental oxygen remains in atomic form in the gas-phase compared with the fraction that is contained in oxygen bearing molecules in the gas or in ices (in the form of $H_2O$, CO, $CO_2$, $H_2CO$, and/or $CH_3OH$). One possible method to address this question is to investigate the isotopic composition of the various molecules present, to determine the relative abundances of the various isotopologues. In dense molecular clouds, photons play a relatively small role and induce very few $^{18}O$ fractionation effects as a result of CO self-shielding (Lyons & Young 2005, Smith et al. 2009). In this case, the most efficient fractionation pathways are exothermic barrierless reactions involving major species ($C^+$, C, O, CO), with the zero-point energy (ZPE) differences in isotopologues driving the fractionation process. Then, the isotopic fractionations are directly linked to the abundance of the elements in the gas-phase when efficient fractionation reactions exist.

Among the various fractionations, hydrogen (deuterium), nitrogen, and carbon have received the most attention (Terzieva & Herbst 2000, Furuya et al. 2011, Pagani et al. 2011, Pagani et al. 2012, Roueff et al. 2015, Furuya & Aikawa 2018), but oxygen fractionation itself has been less well studied. The previous oxygen isotope studies concern mostly the minor isotopologues of CO, ($^{13}CO$, $C^{17}O$, and $C^{18}O$) to probe the molecular content of clouds as the emission lines of $^{12}C^{16}O$ are fully saturated. In addition, $^{13}C^{18}O$ and $^{12}C^{18}O$ have also been used to determine the $^{12}C/^{13}C$ ratio in CO assuming a given $^{16}O/^{18}O$ ratio (Ikeda et al. 2002). The $^{16}O/^{18}O$ ratio across the Galaxy has been derived from the $H_2C^{16}O/H_2C^{18}O$ ratio assuming no fractionation in $H_2CO$ (Wilson 1999). This leads to a local $^{16}O/^{18}O$ ISM value equal to 557 ± 30 (Wilson 1999), close to the Solar System value of 530 for the Solar wind (McKeegan et al. 2011) or 511 ± 10 for the sun's photosphere (Ayres et al. 2013), and around 500 for comets (Bockelée-Morvan et al. 2012, Jehin et al. 2009) and meteorites (Lodders 2003). The $^{16}O/^{18}O$ ratio in diffuse molecular clouds is estimated to be equal to 672 ± 110 from the $HC^{16}O^+/HC^{18}O^+$ ratio (Lucas & Liszt 1998). Apart from $H_2CO$, CO and its protonated form $HCO^+$, there are only a few other scattered detections such as $^{18}$OCS, S$^{18}$O, SO$^{18}$O observations in Orion KL (Tercero et al. 2010, Esplugues et al. 2013), leading to lower apparent $^{16}$O/$^{18}$O ratios than the local ISM value. Despite the oxygen isotope anomalies in solar-system materials (McKeegan et al. 2011), the role of H$_2$CO to determine the $^{16}$O/$^{18}$O ratio and the anomalous results for $^{18}$OCS, S$^{18}$O and SO$^{18}$O in Orion KL, no model has been developed to calculate the oxygen fractionation of these molecules. Indeed, the only $^{18}$O fractionation model, developed by Langer et al. (1984), considers only a few isotopologues and only one fractionation reaction, namely

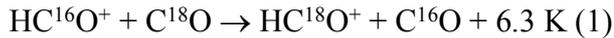
$$HC^{16}O^+ + C^{18}O \rightarrow HC^{18}O^+ + C^{16}O + 6.3 \text{ K} \quad (1)$$

To understand the mechanisms leading to oxygen fractionation, we have developed a new gas-grain model for dense molecular clouds (that do not contain efficient photodissociation processes). After an exhaustive search, we have introduced various oxygen fractionation reactions, which induce notable fractionation effects for molecules such as NO, SO, O$_2$ and SO$_2$. We have also determined the observed isotopic ratios for these species in a number of cold core using existing spectral surveys. To interpret these observations, we have also added reactions for the isotopic fraction of sulphur to the network. The chemical model including the various updates and the model predictions are presented in Section 2. In section 3, the new observed isotopic ratios and their comparison with the model predictions are shown. Our conclusions are presented in Section 4.

## 2 CHEMICAL MODELLING
### 2.1 Model description

Our chemical model is based on Nautilus in its 3-phase version (Ruaud et al. 2016) using kida.uva.2014 (Wakelam et al. 2015), with updates from Ruaud et al. (2015), Wakelam et al. (2017), Hincelin et al. (2015) and Loison et al. (2016), Loison et al. (2017), Vidal et al. (2017) for the chemistry. The network used here is limited to a carbon skeleton up to C$_3$H$_x$N$_y$ (x = 0-2, y=0-1) and C$_3$H$_x$N$_y^+$ (x = 0-3, y=0-1) to reduce the number of reactions when considering all the $^{18}$O and $^{34}$S species. It includes 4440 reactions in the gas phase and 5180 reactions on grains. We have checked that the new network reproduces the abundances of the complete network for the main species studied here. The chemical composition of the gas-phase and the grain surfaces is computed as a function of time. The gas and dust temperatures are equal to 10 K, the H$_2$ density is equal to 2×10$^4$ cm$^{-3}$ (various runs have been performed with the H$_2$ density varied between 1×10$^4$ cm$^{-3}$ and 2×10$^5$ cm$^{-3}$). The temperature and the H$_2$ density is kept constant during chemical cloud evolution. The cosmic-ray ionization rate is equal to 1.3×10$^{-17}$ s$^{-1}$ and the value of the total visual extinction is set to 10. All elements are assumed to be initially in atomic form except for hydrogen, which is entirely molecular. Elements with an ionization potential below the maximum energy of ambient UV photons (13.6 eV, the ionization energy of H atoms), C, S and Fe, are initially in a singly ionized state. The initial abundances are reported in Table 1, the C/O elemental ratio being equal to 0.7 while sulphur is depleted by a factor of 10 (Fuente et al. 2016, Vidal et al. 2017, Fuente et al. 2018). Although this represents a simplistic approach to molecular cloud modeling, neglecting the structure and the history of the cloud, the main objective of this study is to demonstrate the importance of the chemistry on oxygen fractionation. Nevertheless, we performed different runs to test some of the approximations on the fractionation. First, we performed a run with an initial abundance of CO equal to $8.5 \times 10^{-5}$ and an initial abundance of C also equal to $8.5 \times 10^{-5}$ (in the nominal model the initial abundance of CO is equal to 0 and the initial abundance of C is equal to $1.7 \times 10^{-4}$). An initial abundance of CO equal to $8.5 \times 10^{-5}$ represents 35% of the elemental oxygen abundance. The global results of calculations are similar to the nominal model except if we start with an initial $C^{16}O/C^{18}O$ ratio different from the local ISM $^{16}O/^{18}O$ ratio (for example with an initial $C^{16}O$ equal to $8.5 \times 10^{-5}$ but without $C^{18}O$, all the $^{18}O$ being under the atomic form). This may happen if the photodissociation of CO in the first part of cloud evolution (diffuse H I cloud where UV radiation effects are important) leads to complete $C^{18}O$ photodissociation but not for $C^{16}O$ due to self-shielding. Then, the initial $^{16}O/^{18}O$ ratio may be smaller than the local ISM value, with some $^{16}O$ being locked into $C^{16}O$. However, if 35% of elemental oxygen is initially in the form of $C^{16}O$ with no $C^{18}O$ (then with a ratio $^{16}O/^{18}O$ of 320 for atomic oxygen in the gas-phase instead of 500) we cannot reproduce the strong $^{18}O$ enrichment of SO if we do not consider the exchange reactions. Moreover, almost all the oxygenated species, apart from CO and methanol, are moderately enriched in $^{18}O$. We also performed runs varying the total density. The main effect of increasing the total density is to increase the efficiency of the chemistry because the fluxes of the chemical reactions are proportional to the densities of the reactants. However, variation of the total density leads to only minor effects on the chemistry. It should be noted, however, that $S^{18}O$ fractionation is slightly more efficient with a higher total density.

In our model, there are some photons generated by the relaxation of excited $H_2$ (produced by electron collisions with $H_2$) (Prasad & Tarafdar 1983, Gredel et al. 1989). These photons have however only a small effect. Moreover, as we do not consider the photochemical boundary of the molecular cloud, the photodissociation of CO should not play a major role and we do not consider potential self-shielding effects (Lyons & Young 2005, Smith et al. 2009).

**Table 1**. Elemental abundances with respect to hydrogen nuclei.

| Element | Abundance |
|---|---|
| He | 0.09 |
| C | 1.7e-4 |
| N | 6.2e-5 |
| O ($^{16}$O) | 2.4e-4 |
| $^{18}$O | 4.81e-7 |
| $^{16}$O/$^{18}$O | 499 |
| S ($^{32}$S) | 1.5e-6 (factor 10 of depletion) |
| $^{34}$S | 6.67e-7 (factor 10 of depletion) |
| $^{32}$S/$^{34}$S | 22.5 |
| Fe | 1.00e-8 |

## 2.2 $^{18}$O and $^{34}$S exchange reactions

In the interstellar medium, fractionation occurs due to the fact that zero-point energy (ZPE) differences favor the exothermic pathway for barrierless exchange reactions. The rate constants for the fractionation reactions have been studied in detail (Terzieva & Herbst 2000, Roueff et al. 2015). Using the work of Henchman and Paulson (1989), we consider in this work that all reactions involve adduct formation. We also consider only reactions without bimolecular exit channels except for exchange reactions because when there is one, or several exothermic bimolecular exit channels, they are likely to be favored. Then, assuming than the adduct lifetime is long enough to have a statistical energy distribution, we can consider that $k_f$ + $k_r$ = $k_\infty$ ($k_\infty$ is the high-pressure rate constant) (Anderson et al. 1985, Terzieva & Herbst 2000, Roueff et al. 2015).

Then

$k_f = \alpha \times (T/300)^\beta \times f(B,m)/(f(B,m) + \exp(\Delta E/T))$,

$k_r = \alpha \times (T/300)^\beta \times \exp(\Delta E/T)/(f(B,m) + \exp(\Delta E/T))$

with: $k_f$ is the rate constant for the forward reaction, that is reaction towards the right in Table 2, $k_f$ is the rate constant for the reverse reaction, that is reaction towards the left in Table 2, $\alpha$ is in cm$^3$.molecule$^{-1}$.s$^{-1}$, $\beta$ is without units. f(B,m), which depends on the rotational constants (B), masses (m) and symmetries of the reactants (Terzieva & Herbst 2000), is close to 1 except when $O_2$ or $S_2$ is involved in the reaction. $\Delta E$ is the exothermicity of the reaction in Kelvin and is equal to the Zero Point Energy (ZPE) differences. This exothermicity is calculated using vibrational data in the literature when it exists. For some isotopologues, vibrational frequencies are unknown. We then calculated the vibrational frequencies for all isotopologues with Density Functional Theory (DFT) at the M06-2X/AVTZ level using Gaussian 09 software (Frisch et al. 2009) and we scaled the theoretical values to the experimental ones for the main isotopologues.

When no data exist on the association reaction, we use the capture rate constant to determine the value of $k_f + k_r$ for barrierless reactions. It should be noted that this formalism seems to be inappropriate for the $^{18}O + O_2$ reaction (Anderson et al. 1985, Wiegell et al. 1997, Fleurat-Lessard et al. 2003, Rajagopala Rao et al. 2015). For this reaction, there are no experimental data at low temperature and the theoretical studies are not in good agreement with experiment in the 150-350 K range. Then the $^{18}O + O_2$ exchange rate constant is uncertain at low temperature (up to a factor 10), leading to large uncertainties for $O_2$ fractionation only.

Among the oxygen fractionation reactions, two of them play critical roles, $^{18}O + NO$ and $^{18}O + SO$. For the $^{18}O + NO$ reaction there are direct measurements of isotope exchange (Fernando & Smith 1979, Fernando & Smith 1981, Anderson et al. 1985, Cobos & Troe 1985, W. M. Smith 1997), where the rate of exchange is in good agreement with the high pressure rate constant considering an indirect mechanism (Hippler et al. 1975, Fernando & Smith 1979, Baulch et al. 2005). For the $^{18}O + SO$ reaction there are no direct measurements of the isotope exchange reaction but as the association reaction, $O + SO \rightarrow SO_2$ is very exothermic, the lifetime of the excited $SO_2^{**}$ formed should be long enough to apply statistical theory and we use $k_f + k_r = k_\infty$ with $k_\infty$ given by experimental studies (Lu et al. 2003, Cobos et al. 1985). Then for the $^{18}O + SO \rightarrow O + S^{18}O$ rate constant we use the experimental high-pressure value, with $SO_2$ production being negligible at the pressure of dense molecular clouds. The uncertainty of the rate coefficient for the $^{18}O + NO$ and $^{18}O + SO$ exchange reactions should be close to 30% at room temperature and can be as high as a factor 2 at low temperature.

For sulphur fractionation, the $S^+ + SO$ and $S + SO$ reactions may play a role in $^{34}SO$ fractionation. For $S^+ + SO$ we neglect the exchange reaction (exothermic by 8.7 K) as there are exothermic bimolecular exit channels, namely the slightly exothermic $S_2^+ + O$ channel (exothermic by 153 K) and the charge transfer reaction (exothermic by 766 K). When we perform a test considering that only the exchange reaction occurs, we observe no effect on SO fractionation. For $S + SO$ we performed various ab-initio and DFT calculations, namely MRCI(+Q)-F12/AVDZ using Molpro 2012 software (Werner et al. 2012), and RCCSD(T)-F12/AVTZ and M06-2X/AVTZ using Gaussian 09 software, the CCSD(T) and DFT methods being likely only poorly adapted due to the highly multiconfigurational aspect of the $S(^3P) + SO(^3\Sigma^-)$ potential energy surface. At the MRCI level we found no barrier for SSO formation on the ground singlet surface (highly exothermic by 295 kJ/mol) but a barrier equal to 6 kJ/mol for SOS formation (exothermic by only 32 kJ/mol) and a barrier equal to 16 kJ/mol for cyclic-SOS formation (exothermic by 154 kJ/mol). Moreover, we found a barrier for the isomerizations

SSO → OSS and SSO → cyc-SOS, so the exchange reaction is unlikely. However, the surface around the transition states for isomerization is particularly complex, and may also involve the triplet surface, so we cannot completely exclude the possibility of some isomerization. We then performed some runs varying the rate constant for the $^{34}S + SO \rightarrow S + {}^{34}SO$ reaction between zero and the capture rate value. Even for a large rate constant, the effect on $^{34}SO$ fractionation is low due to the low abundance of sulphur atoms in the gas-phase associated with the fact that the exothermicity of the $^{34}S$ exchange process is lower than the thermal energy in dense molecular clouds. As a result, there is little doubt that $^{34}SO$ will show very low enrichment levels.

It should be noted that exchange reactions are efficient only in the gas-phase because on ices we always favor the addition channel (for example $^{18}O + O_2 \rightarrow {}^{18}OOO$ only). However, diffusion and tunneling are mass dependent and are then not strictly equivalent for the various isotopologues, these effects being included in our model but having only a small effect on $^{18}O$ fractionation.

For $O_2$ and $SO_2$ there are two possibilities to have one $^{18}O$. Then the elemental $^{16}O^{16}O/^{16}O^{18}O$ and $S^{16}O^{16}O/S^{16}O^{18}O$ ratios are equal to 250 instead of 500.

**Table 2.** Review of isotopic exchange reactions.
$k_f = \alpha \times (T/300)^\beta \times f(B,m)/(f(B,m) + \exp(\Delta E/T))$ (forward reaction, $C^{18}O + HCO^+ \rightarrow HC^{18}O^+ + CO$)
$k_r = \alpha \times (T/300)^\beta \times \exp(\Delta E/kT)/(f(B,m) + \exp(\Delta E/T))$ (reverse reaction, $HC^{18}O^+ + CO \rightarrow C^{18}O + HCO^+$)

| | Reaction | | α | β | ΔE | f(B,m) | reference |
|---|---|---|---|---|---|---|---|
| 1. | $C^{18}O + HCO^+$ | $\rightarrow HC^{18}O^+ + CO$ | 2.6e-10 | -0.4 | -6.3 | 1 | (Smith & Adams 1980, Mladenovic & Roueff 2014) |
| 2. | $H_3O^+ + H_2^{18}O$ | $\rightarrow H_3^{18}O^+ + H_2O$ | 1.5e-9 | -0.5 | -14.5 | | Proton transfer rate constant similar to the $H_2DO^+$ + $H_2O$ one's (Anicich 2003) |
| 3. | $^{18}O^+ + CO$ | $\rightarrow C^{18}O + O^+$ | 4.4e-10 | 0 | -38.0 | 1 | (Fehsenfeld et al. 1974) |
| 4. | $^{18}O + CO$ | $\rightarrow C^{18}O + O$ | "0" | | -38.0 | | This reaction has a barrier and a negligible rate at low temperature (Inn 1974, Toby et al. 1984, Talbi et al. 2006, Goumans & Andersson 2010) |
| 5. | $^{18}O + SO^+$ | $\rightarrow O + S^{18}O^+$ | 3.0e-10 | 0 | -37.3 | 1 | We assume no barrier for this reaction ($SO^+$ has a reactive doublet ground state and reacts quickly with N atoms (Fehsenfeld & Ferguson 2012)) and a rate constant close to capture rate value. The $S^+ + O_2$ exit channel is slightly endothermic and may play a role (Dotan et al. 1979, Tichý et al. 1979, Smith et al. 1981) |
| 6. | $^{18}O + NO$ | $\rightarrow N^{18}O + O$ | 7.0e-11 | 0 | -36.3 | 1 | (Fernando & Smith 1979, Fernando & Smith 1981, Anderson et al. 1985, Cobos & Troe 1985, W. M. Smith 1997) |
| 7. | $^{18}O + SO$ | $\rightarrow S^{18}O + O$ | 5.3e-11 | 0 | -31.5 | 1 | (Lu et al. 2003, Cobos et al. 1985). We neglect here the radiative association reaction estimated equal to 7.0e-16 at room temperature (Rolfes et al. 1965, Singleton & Cvetanović 1988) |
| 8. | $^{18}O + O_2$ | $\rightarrow O + O^{18}O$ | 3.4e-12 | -0.6 | -32.3 | 2 | (Anderson et al. 1985, Wiegell et al. 1997, Fleurat-Lessard et al. 2003, Rajagopala Rao et al. 2015). The experimental studies cover only the 143 K – 353 K range and are difficult to extrapolate at 10K. The theoretical studies are not in very good agreement with the experimental ones. |
| 9. | $^{18}O + SO_2$ | $\rightarrow O + OS^{18}O$ | "0" | | | | There is a barrier equal to 850 K for this reaction (Naidoo et al. 2005) leading to negligible value at low temperature. |

| 10. | $^{34}S^+ + CS$ | $\to S^+ + C^{34}S$ | 2.0e-9 | -0.4 | -7.4 | 1 | We assume no barrier for this reaction by comparison with $O^+ + CO$, with a rate constant close to capture rate value by comparison with similar $S^+$ reactions (Anicich 2003) |
|---|---|---|---|---|---|---|---|
| 11. | $^{34}S^+ + S_2$ | $\to S^+ + S^{34}S$ | "0" | | -8.7 | | We neglect the exchange reaction favoring the much more exothermic charge transfer (-11650 K). |
| 12. | $^{34}S^+ + SO$ | $\to S^+ + ^{34}SO$ | "0" | | -8.6 | | We neglect the exchange reaction favoring the slightly exothermic $S_2^+ + O$ channel (-153 K) and the charge transfer reaction (-766 K) (see text). |
| 13. | $^{34}S + NS$ | $\to S + N^{34}S$ | 7.0e-11 | 0 | -8.9 | 1 | Same high pressure rate constant than $O + NO$ |
| 14. | $^{34}S + S_2$ | $\to S + S^{34}S$ | 3.0e-11 | 0 | -8.7 | 2 | We assume no dynamical effect such as for $O + O_2$ with a rate constant close to capture. |
| 15. | $^{34}S + SO$ | $\to ^{34}SO + S$ | "0" | | -8.6 | | See text |

## 2.3 Modeling results

The CO, HCO$^+$, OH, O$_2$, OCS, H$_2$CO, CH$_3$OH, NO, SO, and SO$_2$ abundances relative to H$_2$ calculated by our model are shown in Figure 1.

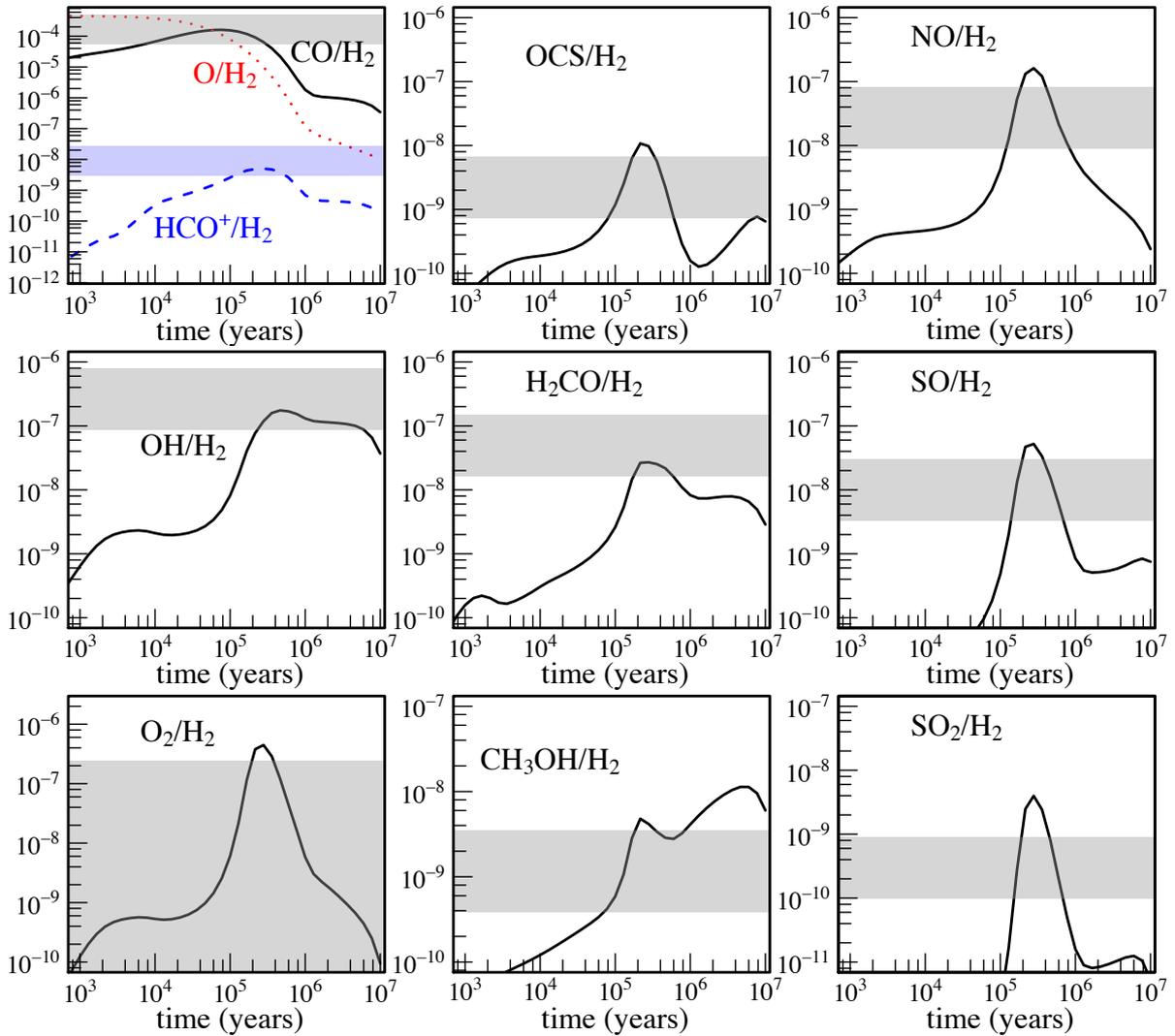

**Figure 1:** Calculated gas-phase abundance ratios, relative to H$_2$, of CO, HCO$^+$, OH, O$_2$, OCS, H$_2$CO, CH$_3$OH, NO, SO and SO$_2$ studied in this work as a function of time predicted by our model for N(H$_2$) = 2×10$^4$ cm$^{-3}$ and T = 10K. The horizontal grey rectangles represent the abundances observed in the cold core TMC-1 (CP),(Gratier et al. 2016, Lique et al. 2006)

including an arbitrary factor 3 for the uncertainties. In this figure, sulphur is depleted by a factor of 10, the S/H$_2$ ratio being equal to $1.5 \times 10^{-6}$.

The calculated $^{16}$O/$^{18}$O ratios for the same species are shown in Figure 2 assuming an elemental $^{16}$O/$^{18}$O ratio of 500.

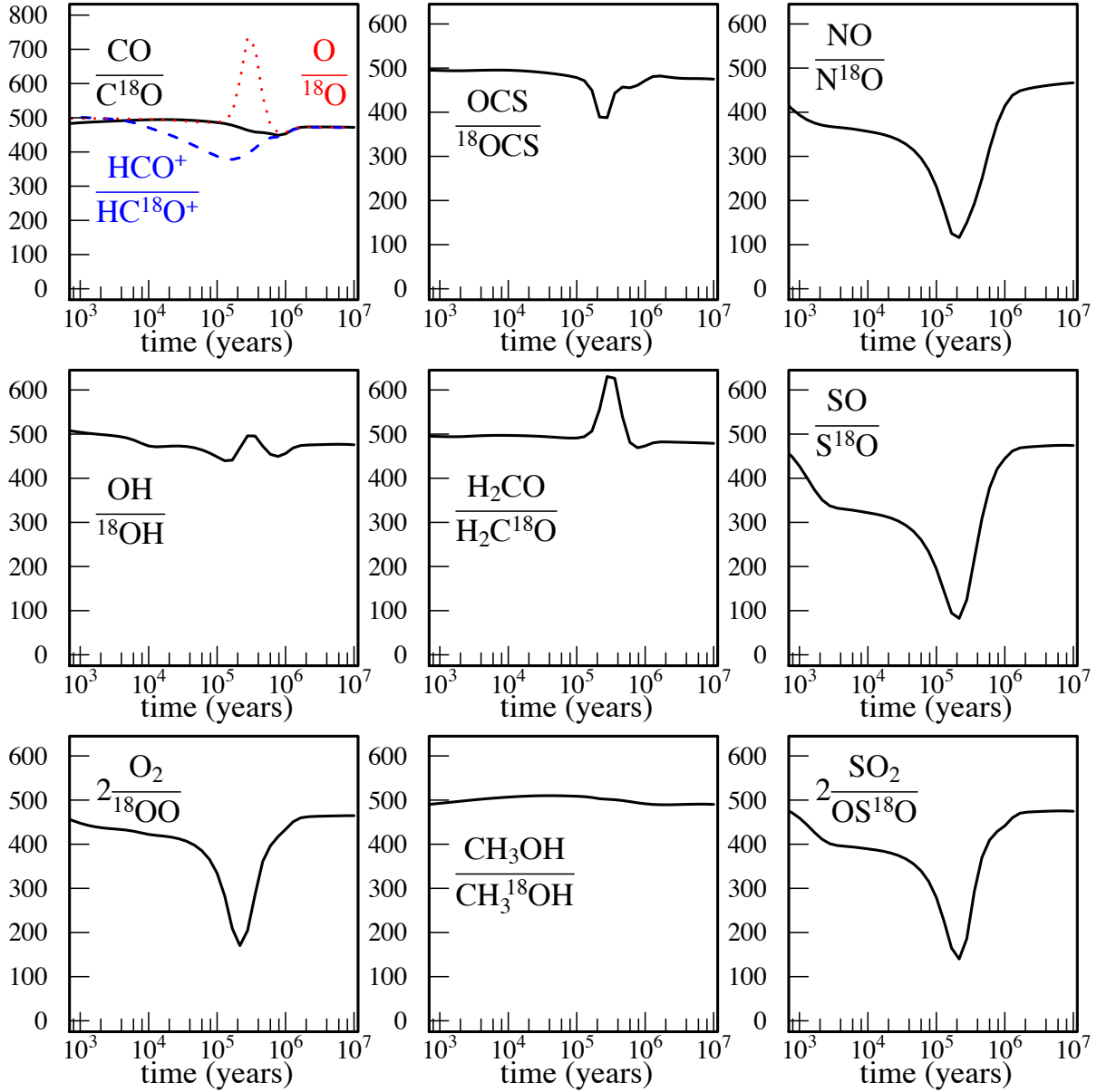

**Figure 2:** Calculated $^{16}$O/$^{18}$O ratio for the main oxygen species in the gas-phase as a function of time predicted by our model (N(H$_2$) = $2 \times 10^4$ cm$^{-3}$, T = 10K). The $^{16}$O/$^{18}$O elemental ratio is taken equal to 500.

The results of our model show variable oxygen fractionation levels with time, some of them being significant. The fractionation for HCO$^+$ induced by reaction (1) is low but non-negligible. As electronic Dissociative Recombination (DR) of HCO$^+$ leads to CO formation, there is some oxygen fractionation in CO too. However, as the DR of HCO$^+$ is not the main source of CO (the main CO sources are the O + CH, O + CH$_2$, O + C$_2$H, O + CN, … reactions), fractionation is less important in CO than in HCO$^+$. The fractionation of methanol is low and similar to the CO one as methanol is essentially produced through CO hydrogenation on ice, where the surface CO, s-CO, comes from depletion of gas-phase CO onto the ice. $^{18}$O enrichment is particularly large for NO, SO, SO$_2$ and O$_2$. This result demonstrates that gas-phase isotopic exchange reactions have a significant effect on the oxygen fractionation in dense interstellar clouds. This effect is important for NO and SO because the fractionation reactions:

$^{18}$O + N$^{16}$O → N$^{18}$O + $^{16}$O + 31.5 K

$^{18}$O + S$^{16}$O → S$^{18}$O + $^{16}$O + 36.3 K

are efficient as the rate constant is close to the capture rate limiting value (see Table 2) and the available exit channels producing N + O$_2$ and S + O$_2$ are endothermic. There is also the equivalent reaction:

$^{18}$O + O$_2$ → $^{18}$OO + $^{16}$O + 32.3 K

which is barrierless but for which the rate constant at low temperature is uncertain due to dynamical effects (see the chemical description above). There is also some fractionation for SO$_2$ and OCS which is brought about by the fractionation of SO, as SO$_2$ is mainly formed through the SO + OH reaction. The fractionation effect on OCS is lower because OCS is mainly formed through the OH + CS reaction (Loison et al. 2012). However, when SO reaches a large abundance, the CH + SO reaction produces large quantities of OCS (Loison et al. 2012), which then partially reflects the SO fractionation level.

The amount of $^{18}$O trapped in CO, SO, NO, O$_2$ and SO$_2$ is large enough around a few 10$^5$ years to strongly deplete gas-phase $^{18}$O. Then, H$_2$CO becomes depleted in $^{18}$O because H$_2$CO is mainly produced through the O + CH$_3$ reaction at this time and atomic oxygen is depleted in $^{18}$O in the gas-phase. At longer times, non-thermal desorption of H$_2$CO produced on ice through CO hydrogenation becomes competitive and H$_2$C$^{18}$O depletion vanishes. Thus, the use of H$_2$CO to determine the O/$^{18}$O ratio across the galaxy (Wilson 1999) may be not the most pertinent choice.

A remarkable point in the predicted $^{18}$O enrichment of SO and NO is its time dependency. Indeed, there is a large variation of the oxygen fractionation as a function of time, with a maximum around 2×10$^5$ years for a constant density of N(H$_2$) = 2×10$^4$ molecules.cm$^{-3}$. After this time, the amount of oxygen ($^{16}$O and $^{18}$O) in the gas-phase strongly decreases where it is either transformed into CO or depleted onto dust grains. Then the $^{18}$O + NO and $^{18}$O + SO reactions are not efficient enough to induce a large fractionation. In contrast to N$^{18}$O, S$^{18}$O is often detected in interstellar media with good sensitivity.

## 3 COMPARISONS WITH OBSERVATIONS

In order to test the results of our chemical model, we analyzed observations from existing surveys obtained with the IRAM 30m telescope in five different cold cores: B1-b, TMC-1, L483, L1689B, and the Horsehead cold dense core. Details on the analysis source by source as well as comparison with our model predictions are given in this section.

### 3.1 Observations

All the observations have been performed with the IRAM 30m telescope. The data were reduced using the CLASS/GILDAS software package, i.e. the individual spectra were co-added, folded to de-convolve from the frequency switching procedure, and a baseline consisting of a low-order polynomial (typically 3 or 5) was withdrawn. The observed lines are presented in Table 3.

**Table 3**: Observed line parameters of SO, $^{34}$SO and S$^{18}$O in B1-b, TMC-1, L483, L1689B and Horsehead cold dense core.

| Molecule | Transition | Frequency (MHz) | $E_{up}$ (K) | $A_{ul}$ (s-1) | VLSR (km.s$^{-1}$) | $D_v$ (km.s$^{-1}$) | $\int T_A^* dv$ (K.km.s$^{-1}$) | τ |
|---|---|---|---|---|---|---|---|---|
| **B1-b** | | | | | | | | |
| SO | 2 2 - 1 1 | 86093.958 | 19.3 | 5.25E-6 | 6.53(2) | 0.74(2) | 1.3714(30) | |
| | | | | | 7.10(2) | 0.59(2) | 0.2158(30) | |
| SO | 2 3 - 1 2 | 99299.886 | 9.2 | 1.12E-5 | 4.90(12) | 0.63(18) | 0.0230(50) | |
| | | | | | 6.42(2) | 0.85(2) | 3.6951(10) | |
| | | | | | 7.14(2) | 0.75(2) | 1.5493(10) | |
| SO | 5 4 - 4 4 | 100029.550 | 38.6 | 1.08E-6 | 6.55(2) | 0.63(2) | 0.0409(10) | |
| SO | 3 2 - 2 1 | 109252.181 | 21.1 | 1.08E-5 | 6.43(4) | 0.58(4) | 0.9196(20) | |
| | | | | | 6.84(4) | 0.59(4) | 0.5749(20) | |
| | | | | | 7.35(13) | 0.68(8) | 0.0774(20) | |
| $^{34}$SO | 2 2 - 1 1 | 84410.685 | 19.2 | 4.95E-6 | 6.57(1) | 0.75(2) | 0.0734(20) | |
| $^{34}$SO | 2 3 - 1 2 | 97715.405 | 9.1 | 1.07E-5 | 6.47(3) | 0.57(5) | 0.5285(50) | |
| | | | | | 6.89(10) | 0.65(5) | 0.3482(50) | |
| $^{34}$SO | 3 2 - 2 1 | 106743.363 | 20.9 | 1.01E-5 | 6.54(2) | 0.60(2) | 0.0779(20) | |
| | | | | | 7.01(2) | 0.31(6) | 0.0079(20) | |
| S$^{18}$O | 2 3 - 1 2 | 93267.376 | 8.7 | 9.34E-6 | 6.59(3) | 0.68(2) | 0.1643(20) | |
| S$^{18}$O | 3 2 - 2 1 | 99803.664 | 20.5 | 8.22E-6 | 6.59(3) | 0.55(8) | 0.0106(10) | |
| | | | | | 7.10(7) | 0.24(10) | 0.0015(10) | |
| **TMC-1** | | | | | | | | |
| SO | 2 2 - 1 1 | 86093.958 | 19.3 | 5.25E-6 | 5.692(5) | 0.24(5) | 0.0048(10) | |
| | | | | | 6.07(1) | 0.35(2) | 0.0146(60) | |
| SO | 2 3 - 1 2 | 99299.886 | 9.2 | 1.12E-5 | 5.668(4) | 0.357(8) | 0.2237(50) | |
| | | | | | 6.082(2) | 0.365(4) | 0.3809(50) | |
| SO | 3 2 - 2 1 | 109252.181 | 21.1 | 1.08E-5 | 5.67(7) | 0.47(14) | 0.0077(20) | |
| | | | | | 6.13(3) | 0.34(6) | 0.0093(20) | |
| $^{34}$SO | 2 3 - 1 2 | 97715.405 | 9.1 | 1.07E-5 | 5.69(2) | 0.29(5) | 0.0110(20) | |
| | | | | | 6.11(2) | 0.35(4) | 0.0219(20) | |
| S$^{18}$O | 2 3 - 1 2 | 93267.376 | 8.7 | 9.34E-6 | 5.78(8) | 0.16(16) | 0.0006(10) | |
| | | | | | 6.11(3) | 0.38(9) | 0.0051(10) | |
| **L483** | | | | | | | | |
| SO | 2 2 - 1 1 | 86093.958 | 19.3 | 5.25E-6 | | | 0.254(25) | |
| SO | 2 3 - 1 2 | 99299.886 | 9.2 | 1.12E-5 | | | 1.61(16) | |
| SO | 5 4 - 4 4 | 100029.550 | 38.6 | 1.08E-6 | 5.59(3) | 0.92(10) | 0.007(1) | |
| SO | 3 2 - 2 1 | 109252.181 | 21.1 | 1.08E-5 | | | 0.226(22) | |
| $^{34}$SO | 2 2 - 1 1 | 84410.685 | 19.2 | 4.95E-6 | 5.29(3) | 0.42(4) | 0.0073(7) | |
| $^{34}$SO | 2 3 - 1 2 | 97715.405 | 9.1 | 1.07E-5 | 5.04(3) | 0.42(3) | 0.173(17) | |
| $^{34}$SO | 3 2 - 2 1 | 106743.363 | 20.9 | 1.01E-5 | 4.92(7) | 0.48(5) | 0.008(1) | |
| S$^{18}$O | 2 3 - 1 2 | 93267.376 | 8.7 | 9.34E-6 | 4.95(3) | 0.43(3) | 0.035(3) | |
| **L1689B** | | | | | | | | |
| SO | 2 2 – 1 1 | 86093.958 | 19.3 | 5.25E-6 | 3.56(0) | 0.50(0) | 0.780(67) | 0.86 |
| SO | 2 3 – 1 2 | 99299.886 | 9.2 | 1.12E-5 | 3.55(0) | 0.65(0) | 2.889(244) | |
| SO | 5 4 - 4 4 | 100029.550 | 38.6 | 1.08E-6 | 3.81(2) | 0.41(5) | 0.009(1) | 0.009 |
| SO | 3 2 - 2 1 | 109252.181 | 21.1 | 1.08E-5 | 3.66(0) | 0.49(0) | 0.758(63) | 0.91 |
| SO | 3 3 – 2 2 | 129138.923 | 25.5 | 2.25E-5 | 3.60(0) | 0.45(0) | 0.671(81) | 1.03 |
| SO | 3 4 – 2 3 | 138178.600 | 15.9 | 3.17E-5 | 3.47(0) | 0.61(0) | 2.156(257) | |
| SO | 4 3 – 3 2 | 158971.811 | 28.7 | 4.23E-5 | 3.54(0) | 0.46(0) | 0.541(62) | 0.82 |
| SO | 4 4 – 3 3 | 172181.403 | 33.8 | 5.83E-5 | 3.53(0) | 0.43(0) | 0.376(42) | 0.56 |
| SO | 5 6 – 4 5 | 219949.442 | 35.0 | 1.34E-4 | 3.61(1) | 0.51(1) | 0.739(95) | 0.99 |
| SO | 1 1 – 0 1 | 286340.152 | 15.2 | 1.40E-5 | 3.51(1) | 0.33(2) | 0.271(37) | 1.20 |
| $^{34}$SO | 2 2 - 1 1 | 84410.685 | 19.2 | 4.95E-6 | 3.56(1) | 0.49(2) | 0.026(3) | 0.03 |
| $^{34}$SO | 2 3 - 1 2 | 97715.405 | 9.1 | 1.07E-5 | 3.29(0) | 0.49(0) | 0.477(48) | 0.64 |
| $^{34}$SO | 3 2 - 2 1 | 106743.363 | 20.9 | 1.01E-5 | 3.19(1) | 0.46(2) | 0.033(4) | 0.04 |
| $^{34}$SO | 3 3 – 2 2 | 126613.93 | 25.3 | 2.12E-5 | 3.23(3) | 0.47(7) | 0.030(6) | 0.04 |
| $^{34}$SO | 4 3 – 3 2 | 155506.795 | 28.4 | 3.96E-5 | 3.50(2) | 0.33(5) | 0.017(3) | 0.04 |
| $^{34}$SO | 4 4 – 3 3 | 168815.135 | 33.4 | 5.50E-5 | 3.55(2) | 0.31(4) | 0.011(2) | 0.03 |
| S$^{18}$O | 2 3 – 1 2 | 93267.270 | 8.7 | 9.34E-6 | 3.22(0) | 0.47(0) | 0.171(17) | 0.18 |
| S$^{18}$O | 3 4 – 2 3 | 129066.190 | 14.9 | 2.58E-5 | 3.54(0) | 0.47(1) | 0.108(2) | 0.12 |
| S$^{18}$O | 4 3 – 3 2 | 145874.490 | 27.5 | 3.27E-5 | 3.60(3) | 0.42(6) | 0.007(1) | 0.009 |
| **Horsehead** | | | | | | | | |
| SO | 2 2 - 1 1 | 86093.958 | 19.3 | 5.25E-6 | 10.55(4) | 0.57(1) | 0.30(1) | 0.18 |
| SO | 2 3 - 1 2 | 99299.886 | 9.2 | 1.12E-5 | 10.55(4) | 0.57(1) | 3.18(1) | 1.39 |
| SO | 3 2 - 2 1 | 109252.222 | 21.1 | 1.08E-5 | 10.55(4) | 0.57(1) | 0.43(1) | 0.21 |
| SO | 3 3 – 2 2 | 129138.923 | 25.5 | 2.25E-5 | 10.55(4) | 0.57(1) | 0.02(1) | 0.28 |
| SO | 3 4 – 2 3 | 138178.56 | 15.9 | 3.17E-5 | 10.55(4) | 0.57(1) | 2.79(1) | 1.37 |
| SO | 4 3 – 3 2 | 158971.811 | 26.7 | 4.23E-5 | 10.55(4) | 0.57(1) | 0.49(3) | 0.27 |
| SO | 3 3 – 3 4 | 201162.805 | 25.5 | 2.75E-5 | 10.55(4) | 0.57(1) | 0.00(1) | 0.00 |
| SO | 2 2 – 2 3 | 210202.256 | 19.3 | 2.84E-5 | 10.55(4) | 0.57(1) | 0.01(2) | 0.00 |
| SO | 2 1 – 1 2 | 236452.293 | 15.8 | 1.42E-6 | 10.55(4) | 0.57(1) | 0.04(1) | 0.01 |
| SO | 3 2 – 2 3 | 246404.588 | 21.1 | 1.00E-6 | 10.55(4) | 0.57(1) | 0.00(1) | 0.01 |
| SO | 4 3 – 3 4 | 267197.756 | 28.7 | 7.11E-7 | 10.55(4) | 0.57(1) | 0.08(2) | 0.00 |
| $^{34}$SO | 2 2 - 1 1 | 84410.685 | 19.3 | 4.95E-6 | 10.55(4) | 0.57(1) | 0.02(1) | 0.01 |
| $^{34}$SO | 2 3 - 1 2 | 97715.405 | 9.1 | 1.07E-5 | 10.55(4) | 0.57(1) | 0.18(1) | 0.05 |
| $^{34}$SO | 3 2 - 2 1 | 106743.363 | 20.9 | 1.01E-5 | 10.55(4) | 0.57(1) | 0.03(1) | 0.01 |
| $^{34}$SO | 3 4 – 2 3 | 135775.728 | 15.6 | 3.00E-5 | 10.55(4) | 0.57(1) | 0.18(1) | 0.05 |
| $^{34}$SO | 4 3 – 3 2 | 155506.795 | 28.4 | 3.96E-5 | 10.55(4) | 0.57(1) | 0.00(2) | 0.01 |
| $^{34}$SO | 3 3 – 3 4 | 202116.596 | 25.3 | 2.78E-8 | 10.55(4) | 0.57(1) | 0.00(2) | 0.00 |
| $^{34}$SO | 2 1 – 1 2 | 237107.767 | 15.8 | 1.41E-6 | 10.55(4) | 0.57(1) | 0.03(1) | 0.00 |
| $^{34}$SO | 3 2 – 2 3 | 246135.724 | 20.9 | 1.01E-6 | 10.55(4) | 0.57(1) | 0.00(1) | 0.00 |
| $^{34}$SO | 4 3 – 3 4 | 265866.874 | 28.4 | 7.13E-7 | 10.55(4) | 0.57(1) | 0.00(1) | 0.00 |
| S$^{18}$O | 2 3 – 1 2 | 93267.376 | 8.7 | 9.34E-6 | 10.55(4) | 0.57(1) | 0.03(1) | 0.01 |
| S$^{18}$O | 3 2 – 2 1 | 99803.664 | 20.5 | 8.22E-6 | 10.55(4) | 0.57(1) | 0.01(1) | 0.00 |
| S$^{18}$O | 3 4 – 2 3 | 129066.19 | 14.9 | 2.58E-5 | 10.55(4) | 0.57(1) | 0.02(1) | 0.01 |
| S$^{18}$O | 4 3 – 3 2 | 145874.49 | 27.5 | 3.27E-5 | 10.55(4) | 0.57(1) | 0.00(1) | 0.00 |
| S$^{18}$O | 2 1 – 1 2 | 239102.492 | 15.7 | 1.37E-6 | 10.55(4) | 0.57(1) | 0.00(1) | 0.00 |
| S$^{18}$O | 3 2 – 2 3 | 245638.781 | 20.5 | 1.00E-6 | 10.55(4) | 0.57(1) | 0.01(1) | 0.00 |
| S$^{18}$O | 4 3 – 3 4 | 262447.093 | 27.5 | 7.13E-7 | 10.55(4) | 0.57(1) | 0.00(2) | 0.00 |

**L483:**

L483 is a cold dense cloud around a Class 0 source, with a column density of H$_2$ of $3 \times 10^{22}$ cm$^{-2}$ (Tafalla et al. 2000). The volume density and kinetic temperature derived by Jorgensen et al. (2002) are $3.4 \times 10^4$ cm$^{-3}$ and 10 K. The observations of L483 were carried out using the IRAM 30m telescope in the framework of molecular line survey in the 3 mm band. Details on these observations can be found in Agúndez et al. (2018b), Agúndez et al. (2018a) and Marcelino et al. (2018) while a thorough description of the observations and the complete survey will be presented by Agúndez et al. (in preparation). Briefly, observations were performed in several sessions from August 2016 to April 2018, with the telescope pointed toward the position of the infrared source IRAS 18148-0440. The main beam of the IRAM 30m telescope at the frequencies observed is in the range 30-21 arcsec. The receiver EMIR E090 was used connected to the FTS backend, which provides a spectral resolution of 50 kHz (0.14-0.18 km s$^{-1}$ at the observed frequencies, which is good enough to resolve lines in L483, which have typical widths of $\approx$ 0.5 km s$^{-1}$). The frequency-switching technique was used to maximize the telescope time. We detect four lines of SO (see Table 3), although only two are taken into account to derive the column density. The line at 99299 MHz is optically thick while that lying at 100029 MHz is only marginally detected. For $^{34}$SO we detect three lines, which allow to constrain the rotational temperature to 4.5 +/- 0.5 K, value that is assumed to hold for all SO isotopologues. Finally, for S$^{18}$O we only detect one line, although with a very good signal-to-noise ratio.

**B1b and TMC-1:**

Barnard 1 (B1) is a dense core with a steep density gradient where $N$(H$_2$) is between $7.6 \times 10^{22}$ cm$^{-2}$ (Daniel et al. 2013) and $1.3 \times 10^{23}$ cm$^{-2}$ (Hirano et al. 1999). The column density of H$_2$ molecules at the position of the cyanopolyyne peak in TMC-1 has been estimated equal to $10^{22}$ cm$^{-2}$ (Cernicharo & Guélin 1987). The B1b and TMC-1 data are part of a 3mm line survey using the IRAM 30-m telescope (see (Cernicharo et al. 2012) and references within). Observations were performed between January and May 2012 towards the positions $\alpha_{J2000}$=03$^h$33$^m$20.8$^s$, $\delta_{J2000}$=31°07'34" in B1-b, and the cyanopolyyne peak in TMC-1 $\alpha_{J2000}$=04$^h$41$^m$41.88$^s$, $\delta_{J2000}$=25°41'27". The EMIR receivers were used connected to the fast

Fourier Transform Spectrometers (FTS) providing a spectral resolution of 50 kHz and covering the 3mm band between 82.5 and 117.5 GHz. At the observed frequencies the velocity resolution ranges from 0.18 km s$^{-1}$ to 0.12 km s$^{-1}$, which is good enough for the typical line widths (~ 0.5 km s$^{-1}$ in TMC-1, and ~ 1 km s$^{-1}$ in B1b), and allows us to resolve the two velocity components in both sources (~ 5.7 and 6.0 km s$^{-1}$ in TMC-1, ~ 6.5 and 7.0 km s$^{-1}$ in B1b, see Table 3). All the observations were performed using Frequency Switching (FSw) mode. Each spectral setup was observed for 2 hrs, resulting in an average rms of 4-6 mK (up to 10-20 mK at the high frequencies close to the end of the band) for both sources. In B1b we detect 4 lines of SO, but the emission is optically thick and we did not compute column densities. We used the 3 lines of $^{34}$SO to perform rotational diagrams and obtain a rotational temperature of 6.5 +/- 0.7 K, a value close to the 8.2 K obtained by Fuente et al. (2016). This temperature was used to compute the column density of S$^{18}$O (only 2 lines observed). SO lines in TMC-1 are optically thin and we obtained a rotational temperature of 3.9 +/- 0.1 K, which was used to obtain column densities for $^{34}$SO and S$^{18}$O since only one line was detected for these isotopologues (see Table 3).

**L1689B:**

The L1689B dense core is located in the ρ-Ophiuchi star forming complex and has a column density equal to $N$(H$_2$) is 1.4 × 10$^{23}$ cm$^{-2}$ (Steinacker et al. 2016). The data presented here were observed during various campaigns spanning 2011-2017 at the IRAM 30-m telescope. The coordinates of the integration position were α$_{J2000}$=16$^h$34$^m$48.3$^s$, δ$_{J2000}$=-24°38'04". The data were taken in frequency switching mode using the EMIR receivers connected to the fast Fourier transform spectrometer (FTS) with a spectral resolution of 50 kHz (at the observed frequencies the velocity resolution is lower than 0.17 km s$^{-1}$, which is good enough to resolve lines in L1689B, which have typical widths of 0.45-0.5 km s$^{-1}$). The average rms reached ranged between 2-4 mK at frequencies below 115 GHz to around 10 mK at 2 mm and 50 mK at 1 mm. The line integrated intensities were converted to the main beam temperature scale. To derive the molecular column, we assumed the same excitation temperatures for all transitions, albeit possibly different to the kinetic temperature. Under this assumption, we calculated the line integrated intensities for a grid of values in excitation temperature and molecular column densities. We then minimized the $\chi^2$ between the modelled line integrated intensities and the observed ones to derive the molecular column densities. One-sigma error bars were determined using the column density interval defined by $\chi^2$min + 2.3, as described in Bacmann and Faure (2016). For the main SO isotopologue, the lines at 99299.886 MHz and at 138178.6 MHz are optically thick with optical depths around 8-15. We have not considered these lines in our analysis.

**Horsehead:**

The dense core of the Horsehead nebula (the shielded core characterized by a large DCO$^+$ abundance J2000 05h 40m 55.73s, −02° 27′ 38″) has total column density estimated to be N(H$_2$) = $3.0 \times 10^{22}$ cm-$^2$ (Gerin et al. 2009). The data is part of the Horsehead WHISPER survey and was observed using position switching at the 30m IRAM telescope. It covers the full 3, 2 and 1mm atmospheric bands with a spectral resolution of 50kHz at 2 and 3mm and 195kHz at 1mm. This corresponds to a velocity resolution between 0.17 km/s and 0.05 km/s depending on frequency, sufficient to well resolve the lines in this source (linewidth of typically 0.45 km/s). The median noises are 8.1 mK, 18.5 mK, and 8.3 mK respectively for the 3, 2 abs 1mm bands. Details about the data reduction can be found in Pety et al. (2012). Radiative transfer calculations were carried out using the LTE approximation as implemented in the Weeds package (Maret et al. 2011). All lines of SO, S$^{18}$O and $^{34}$SO with frequencies inside the observed bands and with upper level energies lower than 30K where selected. A 2.5MHz frequency range was selected around each of these lines and a Bayesian method was used to recover the radiative transfer model parameters. The method is presented in details in Andron et al. (2018). The observed spectra are presented in Figure A1 in the appendix. And the 1D and 2D histogram of the posterior probability distribution function and the comparison of the observations are presented in Figure A2.

In Table 4, we present the observed $^{32}$S$^{16}$O/$^{32}$S$^{18}$O, $^{32}$S$^{16}$O/$^{34}$S$^{16}$O and $^{34}$S$^{16}$O/$^{32}$S$^{18}$O ratios computed in the five cold cores. More details on the observations are given in the appendix. In addition, we report the values observed in the pre-stellar core L1544 by Vastel et al. (2018). Local ISM and Solar system values are also given for comparison.

**Table 4:** Observed $^{32}$S$^{16}$O/$^{32}$S$^{18}$O, $^{32}$S$^{16}$O/$^{34}$S$^{16}$O and $^{34}$S$^{16}$O/$^{32}$S$^{18}$O column density ratios.

|  | N(H$_2$) in cm-$^2$ | $^{32}$S$^{16}$O/$^{32}$S$^{18}$O | $^{32}$S$^{16}$O/$^{34}$S$^{16}$O | $^{34}$S$^{16}$O/$^{32}$S$^{18}$O |
|---|---|---|---|---|
| TMC1 (this work) | $1.0 \times 10^{22}$ | 115 ± 13 | 19.5 ± 2.2 | 5.9 ± 1.4 |
| L483 (this work) | $3.0 \times 10^{22}$ | 158 ± 47 | 31 ± 9 | 5.1 ± 1.5 |
| B1-b (this work) | $7.6 \times 10^{22}$ | - | - | 5.2 ± 1.4 |
| L1689B (this work) | $1.4 \times 10^{23}$ | 70 ± 20 | 23 ± 7 | 3.3 ± 1.1 |
| Horsehead nebula (dense cloud) (this work) | $3.0 \times 10^{22}$ | 170 ± 20 | 27.3 ± 1.2 | 6.1 ± 0.7 |

| | | $^{16}O/^{18}O$ | $^{32}S/^{34}S$ | $(^{16}O\times^{34}S)/(^{18}O\times^{32}S)$ |
|---|---|---|---|---|
| L1544 (Vastel et al. 2018) | $4.5 \times 10^{22}$ | - | - | $4.7 \pm 1.0$ |
| | | | | |
| Local ISM (Wilson 1999) | | $557 \pm 30$ | $\approx 22$ | 25 |
| Solar system (Lodders 2003) | | 500 | 22.5 | 22.2 |

## 3.2 Comparison with the model predictions

In TMC-1, the core region of the Horsehead, L1689B and L483, some SO lines are optically thin allowing a direct determination of the $^{32}S^{16}O/^{32}S^{18}O$ ratio, showing large discrepancies with the elemental ratio. As the $^{32}S^{16}O/^{32}S^{18}O$ ratio predicted by the model is highly dependent on time (see Figures 2 and 3), the observed ratios may then provide strong constraints on the chemical evolution of the cores. We emphasize here that, for simplicity, the physical structure of the cloud remains constant in our simulations. One result, however, remains valid in that for $^{18}O$ fractionation to be efficient, a high gas-phase abundance of $^{18}O$ is required. The agreement between the predicted and the observed $^{32}S^{16}O/^{32}S^{18}O$ ratio in dense molecular clouds requires an atomic oxygen abundance in the gas phase above $1\times10^{-5}$ relative to $H_2$ corresponding to a relatively small time period. Such large gas-phase abundances of neutral oxygen atoms indicate only partial depletion onto ices. There are several approximations in our model (no structure, no description of the first part of cloud evolution where UV radiation is important) that can affect the conclusions. However, considering the large reactive fluxes involved in the exchange reactions, the very good agreement between observations and calculations (for the $S^{18}O$/SO ratio in addition to the abundances of most of the species detected in these molecular clouds) is very unlikely to be coincidental. For all the cold molecular clouds observed in this study, the large $^{18}O$ enrichment factors in SO clearly show that a significant fraction of oxygen is still in the gas-phase, at least for the part of the cloud probed by SO and $S^{18}O$. One another important result of our study is the fact that the similar global molecular abundances and $S^{18}O$ enhancement levels means that the cold molecular clouds observed here are probably at similar stages of chemical evolution with a non-negligible oxygen atom abundance remaining in the gas-phase.

When the observed $S^{16}O$ lines are optically thick, an alternative way to determine the oxygen fractionation level is to analyze the $^{34}S^{16}O/^{32}S^{18}O$ ratios. To study $^{18}O$ and $^{34}S$ fractionation, we compare the various observed $^{34}S^{16}O/^{32}S^{18}O$ ratios with our model predictions including oxygen and sulphur fractionation. As shown in Table 4, the observed $^{34}S^{16}O/^{32}S^{18}O$ ratios are much smaller (around factor of 5) than the local ISM isotope abundance ratios (25). This effect is mainly due to oxygen fractionation as we obtain very low sulphur fractionation as shown in Figure 3.

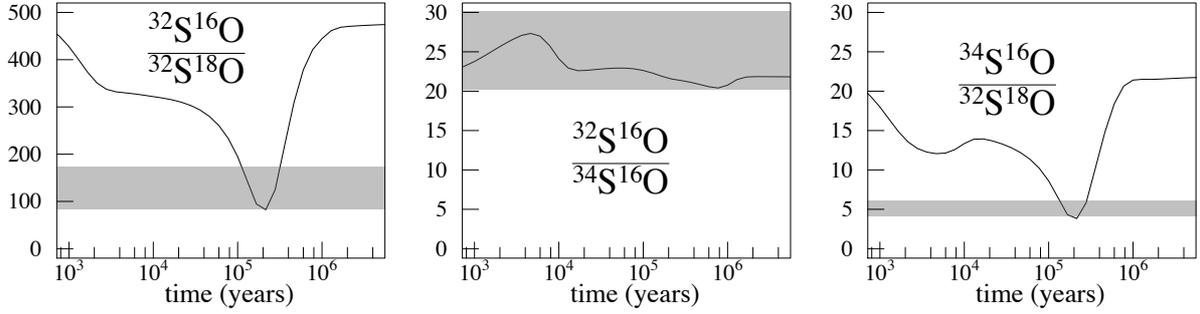

**Figure 3:** Calculated SO/S$^{18}$O, $^{32}$SO/$^{34}$SO and $^{34}$S$^{16}$O/$^{32}$S$^{18}$O ratios as a function of time predicted by our model for n(H) = 2×10$^4$ cm$^{-3}$, T = 10K. The horizontal grey rectangle represents the average of dense core observations listed in Table 4.

Indeed, sulphur fractionation is observed to be very small in SO in the molecular clouds presented here, with a $^{34}$S$^{16}$O/$^{32}$S$^{16}$O ratio close to the local ISM $^{34}$S/$^{32}$S ratio, and also close to the $^{34}$S/$^{32}$S ratio measured in diffuse molecular clouds (Lucas & Liszt 1998) and in the solar system (Lodders 2003). Then $^{34}$S isotopologues are good proxies to derive abundances when the main $^{32}$S species is optically thick. The agreement between the observed and calculated $^{34}$S$^{16}$O/$^{32}$S$^{18}$O ratio leads to the same level of chemical evolution as the one derived using the $^{32}$S$^{16}$O/$^{32}$S$^{18}$O ratio, indicating that a relatively large gas-phase abundance of neutral oxygen atoms must remain to reproduce the observations.

## 4 CONCLUSIONS

In this work, we have presented a gas-grain model for oxygen (and sulphur) isotopic fractionation in cold cores. We also computed observed $^{32}$S$^{16}$O/$^{32}$S$^{18}$O, $^{32}$S$^{16}$O/$^{34}$S$^{16}$O and $^{34}$S$^{16}$O/$^{32}$S$^{18}$O isotopic ratios in five cold cores (B1-b, TMC-1, L483, L1689B, and the Horsehead cold dense core) using existing spectral surveys. Our main result is that the large observed $^{16}$O/$^{18}$O ratio in SO is indicative of large amounts of $^{18}$O in the gas-phase (reactive through neutral-neutral reactions), i.e. moderate depletion of $^{18}$O on grains.

We have also shown that sulphur fractionation is very small in SO and in all species containing sulphur. Consequently, $^{34}$S isotopologues are good proxies to derive abundances when the main $^{32}$S species is optically thick.

The $^{18}$O fractionation levels in SO indicate that this molecule might be a good proxy for leftover material from dense molecular clouds that could be inherited by comets. Indeed, if SO in cometary ices originates from SO formed in the dense core leading to solar system formation, it may show $^{18}$O fractionation. In contrast, if SO is mainly formed from ice photolysis, all $^{18}$O fractionation should disappear as the main reservoir of oxygen on ice, $H_2O$, does not show any $^{18}$O fractionation (Bockelée-Morvan et al. 2012, Mandt et al. 2015). As isotope fractionation is as sensitive probe of the evolution of matter, the chemistry described in this work could eventually be used as the basis for a more complex model including photodissociation and self-shielding effects, allowing us to simulate the evolution of isotopic composition in other environments such as diffuse molecular clouds and planetary systems.

This work was supported by the program "Physique et Chimie du Milieu Interstellaire" (PCMI) funded by CNRS and CNES. VW researches are funded by the ERC Starting Grant (3DICE, grant agreement 336474). Computer time was provided by the Pôle Modélisation HPC facilities of the Institut des Sciences Moléculaires UMR 5255 CNRS − Université de Bordeaux, co-funded by the Nouvelle Aquitaine region. M.A., N.M., J.C. and J.R.C. thank the Spanish MICIU for funding support under grants AYA2016-75066-C2-1-P and AYA2017-85111-P, and the European Research Council for support under grant ERC-2013-Syg-610256-NANOCOSMOS. M.A. also acknowledges funding support from the Ramón y Cajal programme of Spanish MICIU (RyC-2014-16277). IRAM is supported by INSU/CNRS (France), MPG (Germany) and IGN (Spain).

**Appendix:**

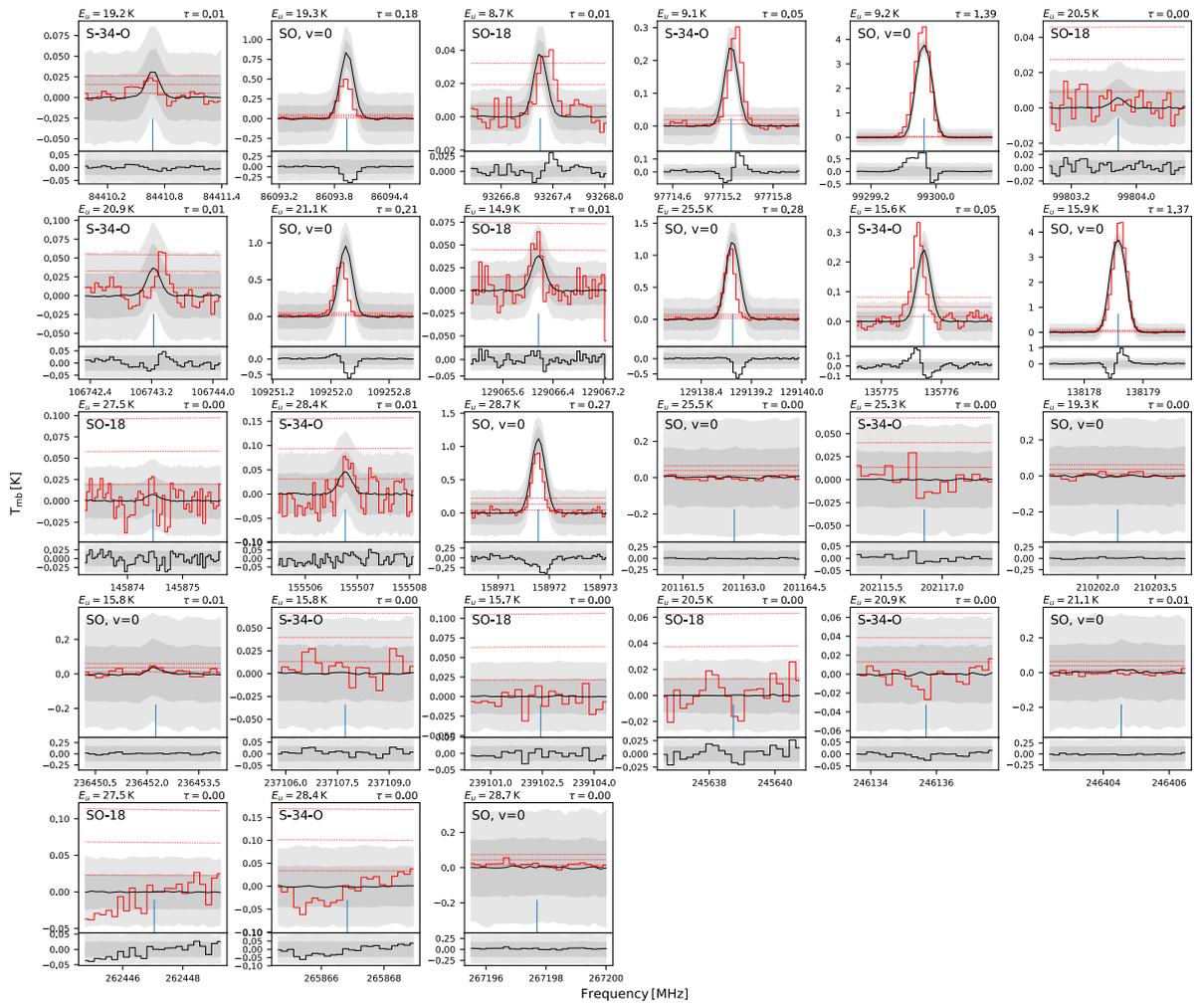

**Figure A1**: Observed spectra (*red line*) of various SO, $^{34}$SO and S$^{18}$O lines at the dense core peak of the Horsehead dense core. The black spectrum is the LTE model for a 8.7 K excitation temperature and a column density of $5.64(0.23) \times 10^{13}$ cm$^{-2}$ for SO, $2.07(0.10) \times 10^{12}$ cm$^{-2}$ for $^{34}$SO and $3.38(0.37) \times 10^{11}$ cm$^{-2}$ for S$^{18}$O. The uncertainties given by the global fitting are shown on grey spectra (68% (1σ) and 95% (2σ) probability).

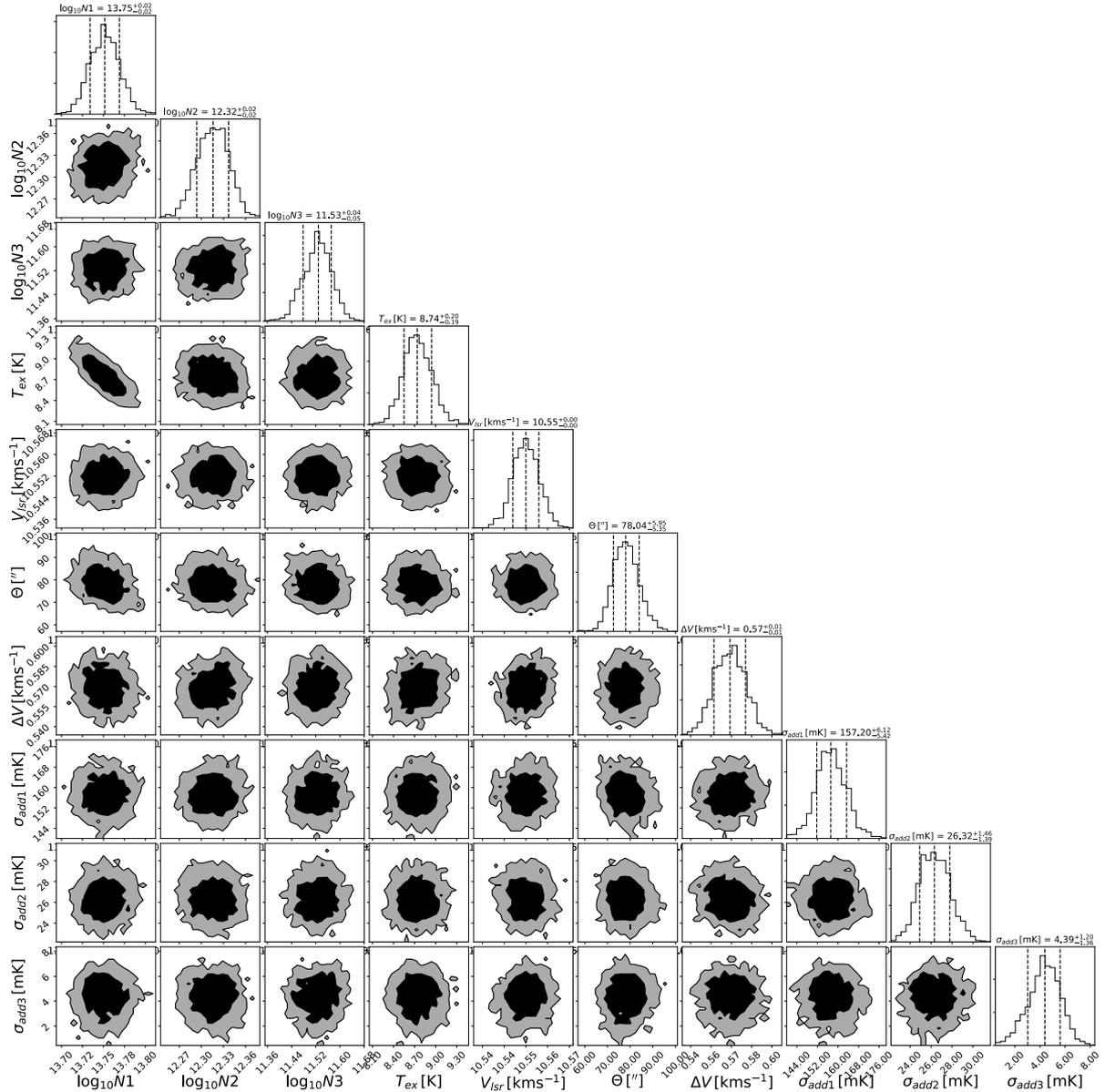

**Figure A2**: Distributions of the posterior probability for SO, $^{34}$SO and S$^{18}$O column densities density (*X1, X2 and X3*), kinetic temperature ($T_{ex}$) (assumed to be equal for all the transition and all the SO isotopologues), $V_{lsr}$ is the systemic velocity assumed to be common to all the species, $\Theta$ is the source size assuming a gaussian source shape centered in the telescope beam, $\Delta V$ is the FWHM of the underlying opacity profile assumed to be common to all the transitions, $\sigma_{add1,2,3}$ is the additional noise added to each species spectra, it is used to take into account possible model discrepancies by increasing the uncertainties on the other parameters in the Horsehead dense core. Along the diagonal, the one-dimensional probability distribution functions are integrations of the two-dimension probability distribution functions displayed below. The color coding of the two-dimensional histograms runs from 0% (white) to 100% of the peak value (black). The grey contour corresponds to 68% (1$\sigma$) of cumulated posterior probability. The Fourier Transform Spectrometer was used with spectral resolutions of 195.3kHz at 1mm and 48.8kHz at 2 and 3mm yielding velocity resolution no higher than 0.17km/s for the lines detected with a peak intensity more than 3 times the noise level. This value is more than 3 times smaller than the estimated linewidth ensuring minimal instrumental broadening.